# Interplay of the pseudo-Raman term and trapping potentials in the nonlinear Schrödinger equation

E. M. Gromov[1, *], B. A. Malomed[2]

[1]National Research University Higher School of Economics, Nizhny Novgorod 603155, Russia

[2]Department of Physical Electronics, School of Electrical Engineering, Faculty of Engineering, and Center for Light-Matter Interaction, Tel Aviv University, Tel Aviv 69978, Israel

**Abstract**

We introduce a nonlinear Schrödinger equation (NLSE) which combines the *pseudo-stimulated-Raman-scattering* (pseudo-SRS) term, i.e., a non-conservative cubic one with the first spatial derivative, and an external potential, which helps to stabilize solitons against the pseudo-SRS effect. Dynamics of solitons is addressed by means of analytical and numerical methods. The quasi-particle approximation (QPA) for the solitons demonstrates that the SRS-induced downshift of the soliton's wavenumber may be compensated by a potential force, producing a stable stationary soliton. Three physically relevant potentials are considered: a harmonic-oscillator (HO) trap, a spatially periodic cosinusoidal potential, and the HO trap subjected to periodic temporal modulation. Both equilibrium positions of trapped pulses (solitons) and their regimes of motion with trapped and free trajectories are accurately predicted by the QPA and corroborated by direct simulations of the underlying NLSE. In the case of the time-modulated HO trap, a parametric resonance is demonstrated, in the form of motion of the driven soliton with an exponentially growing amplitude of oscillations.

**Highlights**
>Solitons in an extended nonlinear Schrödinger equation are studied. >Interplay of a pseudo-Raman-scattering effect and external potentials is considered. >Dynamics of solitons is investigated by means of analytical and numerical methods.

**Keywords**: Extended Nonlinear Schrödinger Equation; Spatial Stimulated Raman Scattering; External Potential; Soliton

## 1. Introduction

Solitons are commonly known as robust localized waves, maintained by the interplay of nonlinearity and dispersion or diffraction in a great variety of physical media [1-4]. Physical settings and the corresponding models, which support soliton modes, may be roughly categorized as high-frequency (HF) and low-frequency (LF) ones. In the former case, solitons appear as spatially broad self-trapped envelopes of HF oscillations, typical examples being optical solitons in Kerr media [1-3] (possibly, also in left-handed metamaterials [5-7]), Langmuir solitons in plasmas [8-10], and nonlinear surface-wave excitations on deep water [11,12]. A universal model for envelope solitons is supplied by the nonlinear Schrödinger equation (NLSE) [13,14]. On the other hand, nonlinear LF waves may build solitons directly in terms of the underlying wave field. Commonly known realizations of the latter type are provided by ion-acoustic [15] and magneto-acoustic [16] waves in plasmas and surface waves on shallow water layers and long internal waves in stratified fluids [17,18], the ubiquitous model being based on the Korteweg – de Vries equation (KdVE) and its generalizations.

---
* Corresponding author. *E-mail address:* egromov@hse.ru (E.M. Gromov).

More complex systems often include wave subsystems of both HF and LF types. Arguably, the most commonly known system featuring nonlinear coupling between HF and LF subsystems models the interaction of Langmuir and ion-acoustic waves in hot plasmas, in the form of the celebrated Zakharov's system (ZS) [11]. Unlike the above-mentioned NLSE and KdVE, ZS is not integrable [19], although its exact single-soliton solutions can be easily found, being stable ones. Other well-known realizations of ZS include the coupling of HF oscillations of atomic positions and long-wave acoustic excitations in long molecular chains [20], as well as the interplay of HF surface and LF internal waves in stratified fluids [21].

While the equation for the LF component in the original ZS admits bidirectional propagation, it can be easily simplified to the unidirectional form, replacing the second-order LF equation by the first-order approximation (it may be, in particular, KdVE) [22], see Eq. (2) below. In the adiabatic approximation, which assumes that the LF component is enslaved to slow evolution of the density of the HF envelope, the former one is eliminated, reducing the ZS to NLSE. On the other hand, a realistic model includes viscosity acting on the LF wave. Straightforward analysis demonstrates that the viscosity gives rise to the first correction to the adiabatic approximation, that still allows one to reduce the underlying ZS to the NLSE, but with an additional nonlinear term including the first-order spatial derivative, which is represented below, in Eq. (3), by coefficient $\mu$. As shown in Refs. [23-25], the latter term, appearing in various settings (see also a short review of previous results in [26]), takes the same form as the well-known stimulated-Raman-scattering (SRS) term in fiber optics [27-29], which has an absolutely different physical nature, being induced by the effect of a small delay (~3 fs) of the nonlinear dielectric response of the silica glass on the temporal-domain evolution of optical pulses [1,3]. In terms of the ZS reduced to the single NLSE, the term under the consideration acts in the spatial domain, and it may be naturally called a *pseudo-SRS* one, there being no real Raman effect in the present setting, while the ZS, which includes the viscosity acting on the LF component, provides effective emulation of optical SRS (note that the SRS terms in optics is a dissipative effect too).

It is well known from fiber optics that the SRS effect leads to the downshift of the soliton's spectrum, and, eventually, to destruction of the self-accelerating narrow solitons [1,3]. Therefore, a relevant issue is development of methods for stabilization of solitons in the presence of SRS, which, as a fundamental effect, is always present in nonlinear fibers. One option is the use of the frequency-sliding optical gain [30]. Other possibilities are the use of the emission of radiation from the soliton's core [31], periodically modulated second-order dispersion [32,33], shifting zero-dispersion point [34], and dispersion-decreasing fibers [35]. In addition to driving the self-acceleration of solitons, the SRS may also drive shock waves in fibers [36,37]

In the NLSE including the pseudo-SRS term, stabilization of solitons is a relevant issue too. In previous works [23-26], a solution was proposed which used variable dispersion, which made it possible to provide stable compensation of the detrimental effect induced by pseudo-SRS.

In this paper, we aim to introduce another setting, based on interplay of the pseudo-SRS term and an external potential, which can help to stabilize and steer solitons in the system. The novelty of the setting is based on the fact that, while the use of the potential is possible (but was not previously considered) in the above-mentioned physical realizations of NLSE derived from the ZS, it could not be implemented in the temporal domain realized in fiber optics (an effective potential cannot be created as a function of the temporal coordinate). We develop an analytical approximation for the analysis of the motion of the soliton, which treats it as a quasi-particle, in combination with systematic simulations of the underlying NLSE. Three types of the external potential, which are occur as ubiquitous ones in various physical setups, are addressed: a harmonic-oscillator (HO) trap, a spatially periodic cosinusoidal potential, and the HO trap under the action of time-periodic modulation, which makes it possible to predict and observe a parametric resonance in the motion of the trapped soliton.

The rest of the paper is organized as follows: the model is formulated, in Section 2, which also includes the derivation of and the quasi-particle approximation (QPA) for the motion of the soliton. This is followed by presentation of results produced by analytical approximations in Section 3. These are various regimes of motion of the soliton under the action of the external potential, obtained in the framework of QPA (including the case of a parametric resonance, produced by a periodically-time-modulated potential), and an analytically calculated correction to the shape of a pinned soliton. Findings generated by direct simulations of the NLSE are summarized, and compared to predictions of the QPA, in Section 4 (subsections 4.1, 4.2, and 4.3 address, severally, the HO, spatially cosinusoidal, and periodically time-modulated OH potentials). The paper is concluded by Section 5.

## 2. The model and quasi-particle approximation

We consider the evolution of slowly varying envelope $U(x,t)$ of the HF wave field (e.g., the local amplitude of Langmuir waves in the plasma) in the cubic medium, under the action of an external potential. The respective NLSE is nonlinearly coupled to an equation for LF variations of the medium's density, $n(x,t)$ (e.g., density of ions the plasma). As a result, we arrive at the ZS-type system [8],

$$2i\frac{\partial U}{\partial t}+\frac{\partial^2 U}{\partial x^2}-2nU-V(x)U=0, \tag{1}$$

$$\frac{\partial n}{\partial t}+\frac{\partial n}{\partial x}-\mu\frac{\partial^2 n}{\partial x^2}=-\frac{\partial |U|^2}{\partial x}, \tag{2}$$

where $\mu$ is the diffusion coefficient in the LF equation. In the lowest adiabatic approximation, Eq. (2) amounts to a local relation, $n=-|U|^2$. Accordingly, Eq. (1) is replaced by the usual NSLE: $2i\partial U/\partial t+\partial^2 U/\partial x^2+2U|U|^2-V(x)U=0$.

In the next approximation, the diffusion term in Eq. (2) is treated as a source of a correction, which yields an amended local relation, $n=-|U|^2-\mu\partial(|U|^2)/\partial x$. The substitution of the latter approximation in Eq. (1) leads to the extended NLSE for the HF amplitude:

$$2i\frac{\partial U}{\partial t}+\frac{\partial^2 U}{\partial x^2}+2U|U|^2+\mu U\frac{\partial(|U|^2)}{\partial x}-V(x)U=0. \tag{3}$$

The term $\sim\mu$ in Eq. (3) represents the above-mentioned pseudo-SRS effect in the spatial domain, which is the crucially important ingredient of the model under the consideration. Its similarity to the SRS effect proper in the temporal domain in optics is seen in the fact that the viscosity term in Eq. (2) may be formally considered as a "spatial delay" of the group-velocity term, $\partial n/\partial x$, which resembles the temporal delay of the nonlinear response of the optical material to the electromagnetic excitation.

With the intention to address localized solutions (quasi-solitons), we adopt zero boundary conditions in the infinite domain, *viz.*, $U|_{x\to\pm\infty}\to 0$. Then, straightforward manipulations with Eq. (3) make it possible to derive the following evolution equations for naturally defined integral moments of the HF field:

$$\frac{dN}{dt}\equiv\frac{d}{dt}\int_{-\infty}^{+\infty}|U|^2 dx=0, \tag{4}$$

$$2\frac{dP}{dt} \equiv 2\frac{d}{dt}\int_{-\infty}^{+\infty} k|U|^2 dx = -\mu\int_{-\infty}^{+\infty}\left[\frac{\partial(|U|^2)}{\partial x}\right]^2 dx - \int_{-\infty}^{+\infty}|U|^2 \frac{dV}{dx}dx, \qquad (5)$$

$$N\frac{d\overline{x}}{dt} \equiv \frac{d}{dt}\int_{-\infty}^{\infty} x|U|^2 dx = \int_{-\infty}^{+\infty} k|U|^2 dx. \qquad (6)$$

Here, we define $U = |U|\exp(i\varphi)$, with $k = \partial\varphi/\partial x$ being the local wavenumber of the wave packet. Obviously, $N$ and $P$ define the total norm and momentum of the wave packet, which is considered as a quasi-particle, $\overline{x}$ being its center-of-mass coordinate.

## 3. Analytical approximations

### 3.1. Effective evolution equations

To analyze of the wave-packet dynamics in the framework of QPA, we assume that the scale of the spatial variation of the external potential in Eq. (3) is much larger than the packet's width, $D_V \gg \Delta$. Relevant solutions to Eq. (3) are then looked for in the form of a sech *ansatz* with constant amplitude $A$ and variable central coordinate $\overline{x}(t)$,

$$U(x,t) = A\operatorname{sech}\left[A(x-\overline{x}(t))\right]\exp\left\{ik(t)x - i(1/2)\int\left[A^2 + V(\overline{x}(t))\right]dt\right\}, \qquad (7)$$

whose norm is $N_{\text{ansatz}} = 2A$. Therefore, the conservation of the total norm as per Eq. (4) implies that $A$ remains a constant, which is corroborated below by direct numerical simulations of Eq. (3). The substitution of *ansatz* (7) in Eqs. (5) and (6), with respect to the condition adopted above, $D_V \gg \Delta \equiv 1/A$, yields what may be considered as equations of motion in the framework of QPA:

$$2\frac{dk}{dt} = -\frac{8}{15}\mu A^4 - \left(\frac{dV}{dx}\right)_{x=\overline{x}(t)}, \qquad \frac{d\overline{x}}{dt} = k. \qquad (8)$$

Equations (8) give rise to a fixed point, i.e., an equilibrium state of the HF field:

$$\left(\frac{dV}{dx}\right)_{x=x_*} = -\frac{8}{15}\mu A^4, \qquad k_* = 0. \qquad (9)$$

The fixed point exists in a region with a negative slope of the external potential, $(dV/dx)_{x=x_*} < 0$. The QPA equations of motion (8) conserve the energy (Hamiltonian), which implies relation

$$\left(\frac{d\overline{x}}{dt}\right)^2 + \frac{8}{15}\mu A^4 \overline{x} + V(\overline{x}) = H, \qquad (10)$$

where $H$ is a constant.

**3.1.1. For the HO potential profile**, $V(x) = px^2$ (with $p > 0$), Eq. (10) with constant $A^2$, i.e., a constant effective driving force, amounts to the energy-conservation equation for the HO with a shifted equilibrium position:

$$\left(\frac{d\overline{x}}{dt}\right)^2 + p\left(\overline{x} + \frac{4}{15}\frac{\mu A^4}{p}\right)^2 = \tilde{H} \geq 0, \qquad (11)$$

where $\tilde{H}$ is another constant, and the respective fixed point $\left(d\bar{x}/dt=0,\ \bar{x}_*=-(4/15)\mu A^4/p\right)$ corresponds to $\tilde{H}=0$. For $\tilde{H}>0$, Eq. (11) gives rise to elliptic trajectories in the plane of $(d\bar{x}/dt,\bar{x})$. The accuracy of this prediction is verified below, in subsection 4.1, by comparing a numerically found frequency of the shuttle motion of the trapped soliton with its counterpart predicted by Eq. (11).

**3.1.2. For a spatially periodic potential profile**, $V(x)=-V_0\cos(2\pi x/L)$ (with $V_0>0$), Eq. (10) with rescaled variables,

$$\rho \equiv \bar{x}/L,\quad \tau \equiv 2tA^2\sqrt{\frac{2}{15}\frac{\mu}{L}}, \tag{12}$$

amounts to the energy-conservation equation:

$$\left(\frac{d\rho}{d\tau}\right)^2 + \rho - a\cos(2\pi\rho) = H, \tag{13}$$

where $a \equiv \dfrac{15}{8}\dfrac{V_0}{\mu L A^4} > 0$. Trajectories defined by Eq. (13) are displayed in Fig. 1 for $a=1$ and $H=0$. Three types of solutions can be identified in this case: a quiescent soliton pinned to the fixed point $(d\rho/d\tau=0,\ \rho=1)$, a trapped soliton periodically moving in a local potential well (curve 1), and a free soliton, moving with an average acceleration above the trapping potential (curve 2, in which the initial positive velocity changes its sign under the action of the constant negative driving force in Eq. (8)).

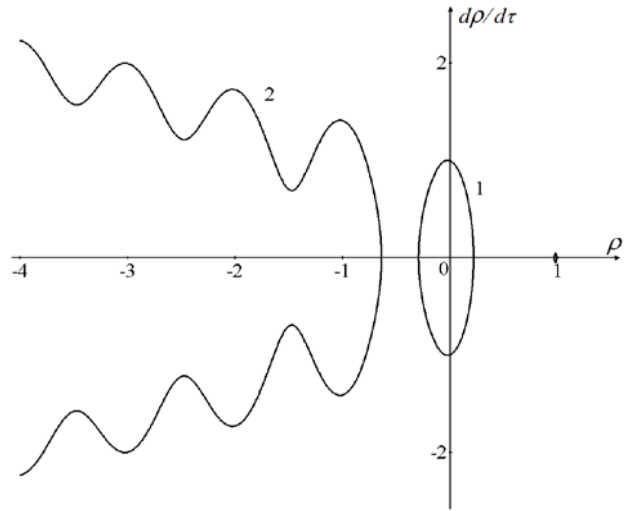

Fig. 1. Quasi-particle trajectories determined by Eq. (13) for $a=1$ and $H=0$. Three types of solutions are shown here: the soliton pinned to fixed point $(d\rho/d\tau=0,\ \rho=1)$, a trapped soliton periodically moving in a local potential well (curve 1), and a free soliton moving with acceleration above the trapping potential (curve 2).

**3.1.3. For the periodically time-modulated HO potential**, *viz.*,
$$V(x,t) = [p_0 + p_1\cos(\Omega t)]x^2, \tag{14}$$
with $p_0 > p_1 > 0$, a parametric resonance in the shuttle motion of a trapped soliton, with frequency $\omega$, is expected when the modulation frequency $\Omega$ in Eq. (14) is close (or exactly equal) to $2\omega$ [38].

In particular, for $p_0 = 1/100$ the frequency of the shuttle motion is $\omega = 1/10$, hence the parametric-resonance frequency is expected to be $\Omega_R = 1/5$. Figure 2 displays a typical solution of system (8) for potential (14) with $p_0 = 1/100$, $p_1 = 1/110$, and $\mu = 1/20$, in the case of the exact parametric resonance, $\Omega = 1/5$. The exponential growth of the amplitude is clearly seen, in agreement with the theory [38]. On the other hand, in the non-resonant case, such as one with $\Omega = 1 \neq \Omega_R = 1/5$, the soliton's parameters vary in time periodically (Fig. 3).

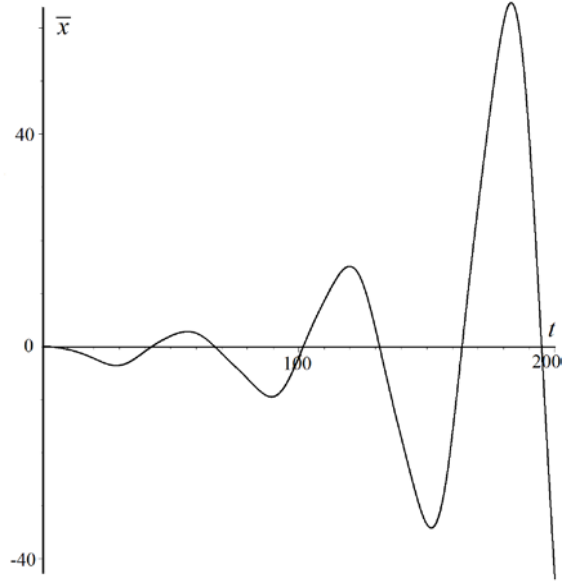

Fig.2. A numerical solution of the quasi-particle equation of motion (8) with potential (14) for the case of the exact parametric resonance case, $\Omega = \Omega_R = 1/5$ (see details in the text).

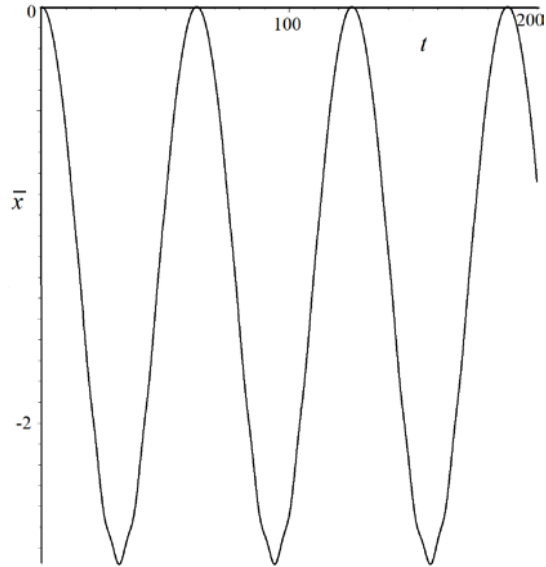

Fig. 3. A numerical solution of the quasi-particle equation of motion (8) with potential (14) far from the parametric resonance, *viz.*, at $\Omega = 1 \neq \Omega_R = 1/5$.

## 3.2. A shape correction to the pinned soliton

Here we address solutions of Eq. (3) for the soliton pinned to the fixed point. The solution is taken as $U(x,t) \equiv \psi(x)\exp[i(\Omega + V(x_*)/2)t]$, for the external potential approximated by the linear profile (in other words, for a constant driving force applied by the potential), in the absence of time modulation: $V(x) = V(x_*) + (dV/dx)_{x_*}(x - x_*)$:

$$-2\Omega\psi + \frac{d^2\psi}{dx^2} + 2\psi^3 + \mu\psi\frac{d(\psi^2)}{dx} - \left(\frac{dV}{dx}\right)_{x_*}(x - x_*)\psi = 0, \quad (15)$$

where $x_*$ is the fixed point given by Eq. (9). Similar to what was adopted above, the linear profile of the potential implies that the wave-packet's width is much smaller than the scale imposed by the potential, $\Delta \ll D_V$. Thus, defining the small parameter, $\varepsilon \sim \Delta/D_V \sim \mu \ll 1$, a solution to Eq. (15) can be looked for as $\psi = \Phi + \phi$, where $\phi$ is a correction $\sim \varepsilon$, which is generated by the last term in the equation. Separating terms of orders $\varepsilon^0$ and $\varepsilon^1$, we obtain

$$\frac{d^2\Phi}{d\xi^2} + 2\Phi^3 - 2\Omega\Phi = 0, \quad (16)$$

$$\frac{d^2\phi}{d\xi^2} + [6\Phi^2 - 2\Omega]\phi = \left(\frac{dV}{dx}\right)_{x_*}\Phi\xi - \frac{2}{3}\mu\frac{d(\Phi^3)}{d\xi}, \quad (17)$$

where $\xi \equiv x - x_*$. Equation (16) has the standard sech-soliton solution, $\Phi = A\,\mathrm{sech}(A\xi)$, where $2\Omega = A^2$. In terms of rescaled variables, $\eta \equiv A\xi$ and $\phi = \Psi V'_\eta A$, Eq. (17) takes the form of a linear inhomogeneous equation,

$$\frac{d^2\Psi}{d\eta^2} + \left(\frac{6}{\cosh^2\eta} - 1\right)\Psi = \frac{\eta}{\cosh\eta} + \frac{2\mu A^3}{V'_\eta}\frac{\sinh\eta}{\cosh^4\eta}. \quad (18)$$

An essential result is that, at

$$V'_\eta = -8\mu A^3/15, \quad (19)$$

which is tantamount to Eq. (9), Eq. (18) has an *exact* localized solution for the correction to the standard sech soliton,

$$\Psi(\eta) = (1/4)(\mathrm{sech}\,\eta)\left[2\eta - \eta^2\tanh\eta - 3(\tanh\eta)\ln(\cosh\eta)\right]. \quad (20)$$

Finally, in terms of the original notation, the corrected soliton solution to Eq. (3) is written as

$$U(x,t) = \frac{A}{\cosh(A\xi)}\left[1 + \frac{(V'_x)_{x_*}}{4}\left[2A\xi - (A\xi)^2\tanh(A\xi) - 3(\tanh(A\xi))\ln(\cosh(A\xi))\right]\right]$$
$$\times \exp\left[\frac{i}{2}(A^2 + V(x_*))t\right]. \quad (21)$$

The accuracy of the analytical correction is verified below (in subsection 4.1) by comparison with full numerical results.

## 4. Results of full numerical simulations

To check the above analytical predictions, we here aim to simulate the evolution of the initial wave packet, $U(x,0) = \mathrm{sech}(x - x_0)$, in the framework of Eq. (2) with different permanent and time-modulated profiles of the external potential and different initial positions $x_0$.

**4.1. For the HO potential**, $V = x^2/100$, the strength of the pseudo-SMS effect is set to be $\mu = 1/10$, and different values of $x_0$ are considered. The respective point (9) of the equilibrium between the pseudo-SRS and potential force is $x_* = -8/3$. In the simulations performed with $x_0 = x_* \equiv -8/3$, at times $t > 10$ the pulse evolves into a stationary localized profile with zero wavenumber. Figure 4 shows the deviation of the absolute value of the numerically found stationary profile from the sech-shaped input, i.e., $\phi_{num}(\xi) = |U(\xi)| - \text{sech}(\xi)$, (the solid curve in the figure), where $\xi \equiv x - x_*$. As shown by the dashed curve in Fig. 2, the deviation is very close to the analytically predicted correction (21), which takes the following form for the current values of the parameters:

$$\Psi(\eta) = -(1/75)(\text{sech}\,\xi)\left[2\xi - \xi^2 \tanh\xi - 3(\tanh\xi)\ln(\cosh\xi)\right]. \tag{22}$$

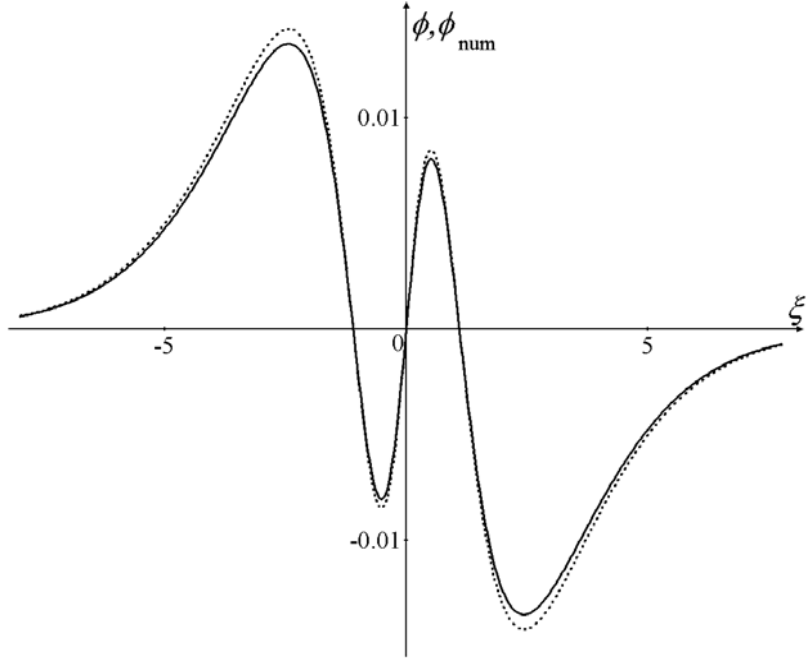

Fig. 4. Numerical results: the deviation of the absolute value of the numerically found stationary pulse from the sech *ansatz*, $\phi_{num}(\xi) = |U(\xi)| - \text{sech}(\xi)$ (the solid curve), with $\xi \equiv x - x_*$. The analytical correction $\phi$ to the absolute value of the stationary solution, given by Eq. (21), is shown by the dashed curve.

A change of the initial position, $x_0$, leads to nonstationary solutions. In particular, Fig. 5 shows the simulated spatiotemporal evolution of $|U(x,t)|$ for $x_0 = 0 \neq x_*$. In this case, the soliton performs robust oscillations with period $T_{num} = 63$ without any visible radiation loss (the latter fact may be explained by the action of the trapping HO potential on the radiation field). The oscillation period predicted by Eq. (8) (or Eq. (11)) for these parameters is $T = 20\pi \approx 62.8$, which is very close to the value revealed by the direct simulations.

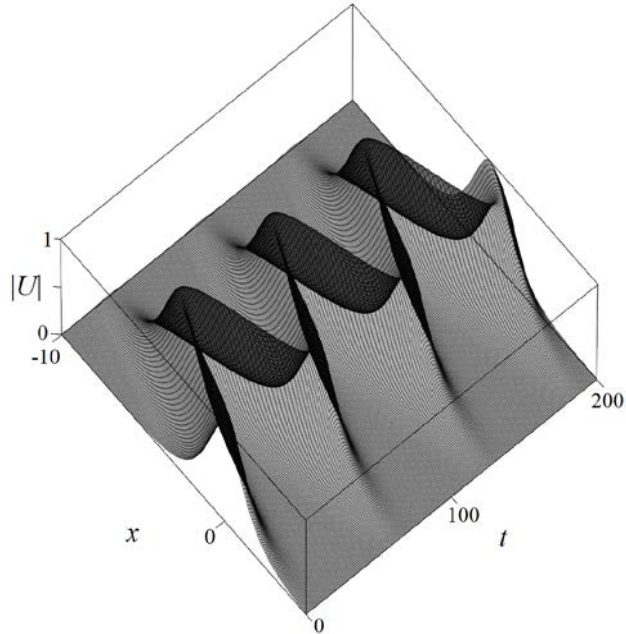

Fig. 5. Results of the simulations of the evolution of the sech-shaped pulse for $x_0 = 0 \neq x_*$ in the HO trapping potential.

**4.2. For the above-mentioned spatially periodic potential profile**,

$$V(x) = -V_0 \cos(2\pi x / L), \qquad (23)$$

we report typical numerical results, taking $L = 40$, $V_0 = 1$, $\mu = 1/20$ and different values of $x_0$ for the initial pulse. Similar to the predictions of QPA, the direct simulations produce solutions of three types: a stable stationary pulse pinned to the equilibrium position, for $x_0 = x_* = 40$ (Fig. 6), which corresponds to point $(d\rho/d\tau = 0, \rho = 1)$ in Fig. 1; a trapped pulse periodically moving in a local potential well (corresponding to curve 1 in Fig. 1) for $x_0 = 0$ (Fig. 7); and a free soliton moving with acceleration above the trapping potential for $x_0 = -20$ (Fig. 8), corresponding to curve 2 in Fig. 1. In particular, the numerically found period of the shuttle motion of the trapped soliton in Fig. 7 is $T_{\text{num}} \approx 56.5$, while QPA predicts, by means of Eq. (8), $T \approx 57$ for the same parameters, which once again corroborates the accuracy of QPA.

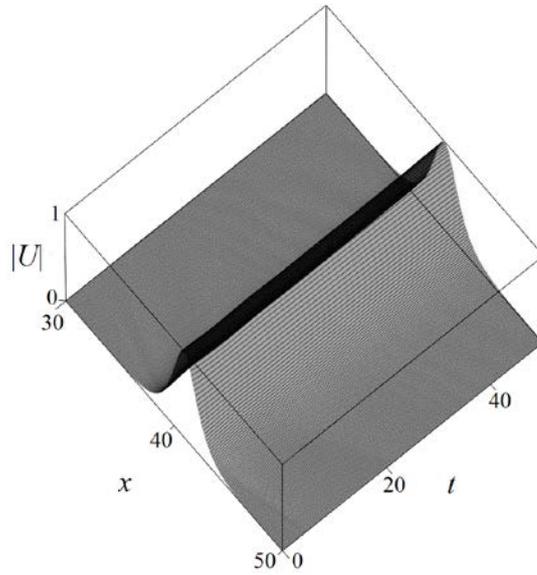

Fig. 6. Results of the simulations of Eq. (3) for the evolution of the sech-shaped pulse with the spatially periodic potential (23) for $x_0 = x_* = 40$.

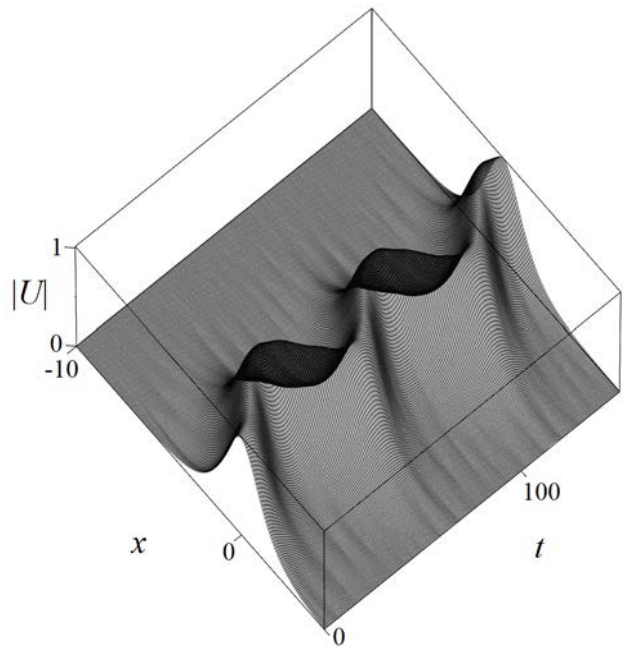

Fig. 7. Results of the simulations of Eq. (3) for the shuttle motion of the sech-shaped pulse in the spatially periodic potential (23) for $x_0 = 0 \neq x_*$; cf. Fig. 5, which displays a similar result obtained under the action of the HO potential.

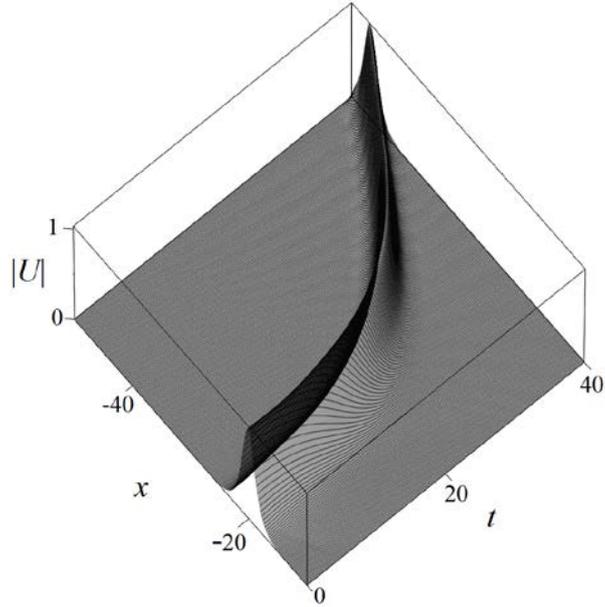

Fig.8. Results of the simulations of Eq. (3) for the evolution of the sech-shaped pulse in the spatially periodic potential (23) for $x_0 = -20$: accelerating motion of a free soliton above the trapping potential.

**4.3. For the periodically time-modulated HO potential** (14), the frequency of the shuttle motion is $\omega = \sqrt{p_0}$, for instance, $\omega = 1/10$ for $p_0 = 1/100$. As said above, the exact parametric resonance corresponds to the time-modulation frequency $\Omega_R = 2\omega = 1/5$. Figure 9 displays results of the simulations of the evolution of the sech-shaped pulse under the action of the time-modulated potential with $\Omega = \Omega_R = 1/5$, $p_1 = 1/110$, $x_0 = 0$, and $\mu = 1/20$. The results clearly agree with the resonance-induced instability, predicted by the numerical solution of the QPA equation of motion (8) in Fig. 2. Simulations for a typical non-resonant case, with $\Omega = 1 \neq \Omega_R = 1/5$, is displayed in Fig. 10, which shows agreement with its counterpart produced by Eq. (8), see Fig. 3.

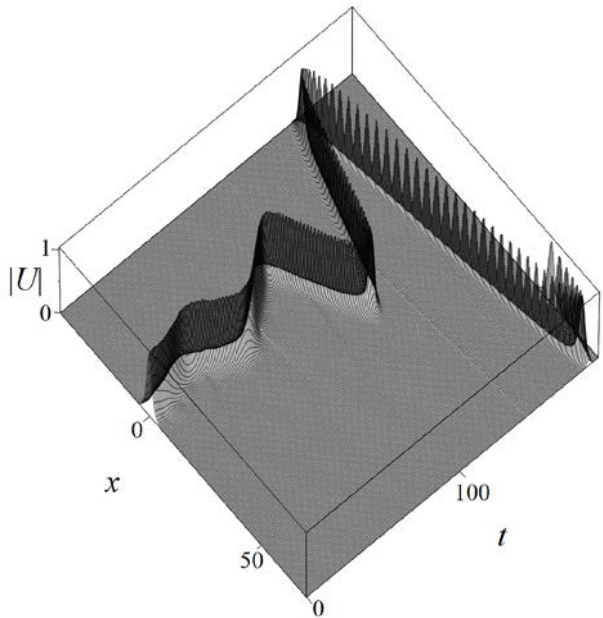

Fig. 9. Results of the simulations of Eq. (3) for the evolution of the sech-shaped pulse in the periodically time-modulated potential (14) with $\Omega = \Omega_R = 1/5$ (the case of the exact parametric resonance, cf. Fig. 2).

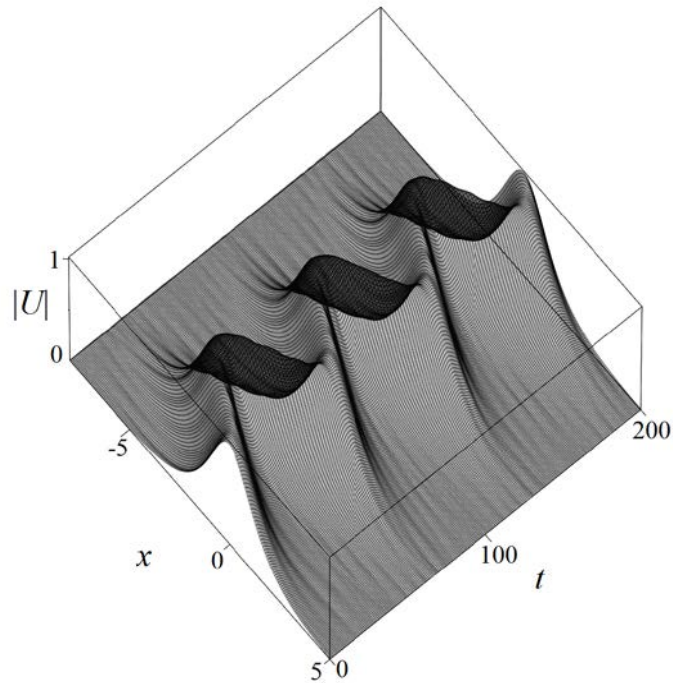

Fig.10. Results of the simulations of Eq. (3) for the evolution of the sech-shaped pulse in potential (14) with $\Omega = 1 \neq \Omega_R = 1/5$, which is taken far from the resonance, cf. Fig. 3.

## 5. Conclusions

We have considered the dynamics of solitons in the framework of the extended NLSE, which includes the pseudo-SRS (simulated Raman scattering) term and an external potential. The NLSE can be derived from the ZS (Zakharov system) whose HF and LF (high-frequency and low-frequency) components are subject, respectively, to the action of the external potential and effective diffusion. The model can be realized in various physical settings, such as hot plasmas, coupled surface and internal waves in stratified fluids, etc. In the framework of the extended NLSE, the interplay of the effective driving force, induced by the pseudo-SRS term, and the potential force creates an equilibrium position of the soliton, and gives rises to various dynamical regimes. Three physically relevant potentials are considered in detail: a static HO (harmonic-oscillator) trap, a spatially periodic potential (23), and the HO trap subject to the time-periodic modulation, as defined by Eq. (14). In the case of the HO potential, the location of the equilibrium position and frequency of oscillations around it are accurately predicted by the QPA (quasi-particle approximation), which treats the soliton as a mechanical particle subject to the action of the two aforementioned forces. In the presence of the periodic time modulation applied to the HO trap, the parametric resonance is predicted by the QPA, and confirmed by direct simulations. Under the action of the spatially periodic potential, the equilibrium position of the soliton, as well as its trapped and free trajectories, are found too by means of both the QPA and full simulations.

## Acknowledgments

The work of BAM is supported, in part, by the Israel Science Foundation via grant No. 1286/17.